\documentclass[12pt]{article}
\usepackage{graphicx}
\usepackage{amssymb}

\newcommand\pubnumber{DPF2015-43}
\newcommand\pubdate{\today}

\def\napoli{Physik Institut\\
University of Zurich, Switzerland}
\def\support{\footnote{On the behalf of the LHCb collaboration.}}

\def\Title#1{\begin{center} {\Large #1 } \end{center}}
\def\Author#1{\begin{center}{ \sc #1} \end{center}}
\def\Address#1{\begin{center}{ \it #1} \end{center}}

\newcommand\pubblock{\rightline{\begin{tabular}{l} \pubnumber\\
         \pubdate  \end{tabular}}}
\newenvironment{Abstract}{\begin{quotation}  }{\end{quotation}}
\newenvironment{Presented}{\begin{quotation} \begin{center} 
             PRESENTED AT\end{center}\bigskip 
      \begin{center}\begin{large}}{\end{large}\end{center} \end{quotation}}

\def\beq{\begin{equation}}
\def\eeq#1{\label{#1}\end{equation}}
\def\eeqn{\end{equation}}

\def\beqa{\begin{eqnarray}}
\def\eeqa#1{\label{#1}\end{eqnarray}}
\def\eeqan{\end{eqnarray}}

\let\bar=\overbar

\def\Dslash{\not{\hbox{\kern-4pt $D$}}}
\def\dslash{\not{\hbox{\kern-2pt $\del$}}}

\def\msb{{\bar{\ssstyle M \kern -1pt S}}}

\begin{document}
\begin{titlepage}
\pubblock

\vfill
\Title{Search for a light dark sector particle at LHCb}
\vfill
\Author{ Andrea Mauri\support}
\Address{\napoli}
\vfill
\begin{Abstract}
A search is presented for a hidden-sector boson, $\chi$, produced in the decay $B^0 \rightarrow K^* (892)^0 \chi$, with $K^* (892)^0 \rightarrow K^+ \pi^-$ and $\chi \rightarrow \mu^+ \mu^-$ . The search is performed using a $pp$-collision data sample collected at $\sqrt{s}=7$ and 8 TeV with the LHCb detector, corresponding to integrated luminosities of 1 and 2 fb$^{-1}$ respectively.
No significant signal is observed in the mass range $214 \le m_\chi \le 4350$ MeV, and upper limits are placed on the branching fraction product $\mathcal{B}(B^0 \rightarrow K^* (892)^0 \chi) \times \mathcal{B}(\chi \rightarrow \mu^+ \mu^- )$ as a function of the mass and lifetime of the $\chi$ boson. These limits place the most stringent constraints to date on many theories that predict the existence of additional low-mass dark bosons.

\end{Abstract}
\vfill
\begin{Presented}
DPF 2015\\
The Meeting of the American Physical Society\\
Division of Particles and Fields\\
Ann Arbor, Michigan, August 4--8, 2015\\
\end{Presented}
\vfill
\end{titlepage}
\def\thefootnote{\fnsymbol{footnote}}
\setcounter{footnote}{0}

\section{Introduction}

Most extensions of the Standard Model (SM) that address the problem of the existence of Dark Matter, postulate the existence of a hidden sector, see for example the review in Ref. \cite{Dark_sector}.
Particles of the hidden sector are singlets with respect to the SM gauge number, however they can interact with SM particles via kinetic mixing. In this analysis a search for a light scalar particle (dark scalar boson, $\chi$) belonging to the secluded sector and mixing with Higgs boson is performed. 
Concrete examples of such models are theories where such a $\chi$ field was responsible for an inflationary period in the early universe \cite{Bezrukov:light_infl_hunters_guide}, and the associated inflaton particle is expected to have a mass in the range $270 < m(\chi) < 1800$ MeV.
Another class of models invokes the axial-vector portal \cite{Axion} in theories of dark matter that seek to address the cosmic-ray anomalies, and to explain the suppression of charge-parity (CP) violation in strong interactions \cite{Peccei:axionCP}.
These theories postulate an additional fundamental symmetry, the spontaneous breaking of which results in a particle called the axion \cite{Peccei:axion}. The energy scale, $f(\chi)$, at which the symmetry is broken lies in the range $1 \lesssim f(\chi) \lesssim 3$ TeV \cite{Nomura:2008ru}.

\begin{figure}[b]
	\centering
	\includegraphics[width=8cm]{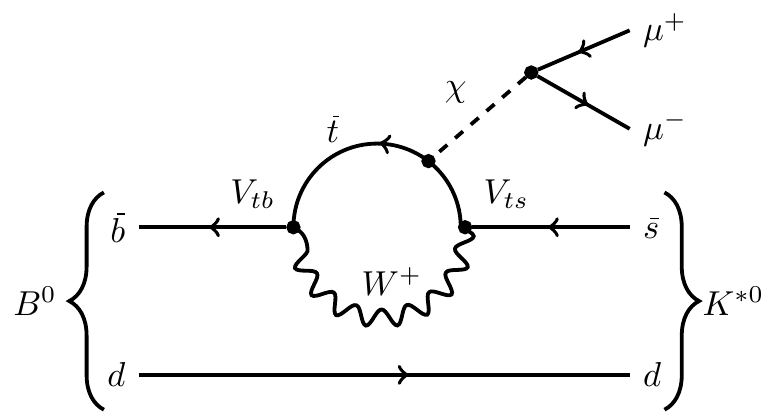}
	\caption{Feynman diagram for the decay $B^0 \rightarrow K^{*0} \chi$, with $\chi \rightarrow \mu^+ \mu^-$.}
	\label{fig:Feymn_diag}
\end{figure}

\section{Search for $B^0 \rightarrow K^* (892)^0 \chi (\rightarrow \mu^+ \mu^-)$}

The decay $B^0 \rightarrow K^{*0} \chi$, with $K^{*0} \rightarrow K^+ \pi^-$ and $\chi \rightarrow \mu^+ \mu^-$ is studied to search for such a hidden-sector particle. 
An enhanced sensitivity to hidden-sector bosons arises because the $b \rightarrow s$ transition is mediated by a top quark loop at leading
order (Fig.\ref{fig:Feymn_diag}). Therefore, a $\chi$ boson with $2m(\mu) < m(\chi) < m(B^0) - m(K^{*0})$ and a sizable top quark coupling (obtained via mixing with the Higgs sector), could be produced at a substantial rate in such decays.

Similar searches have been performed in the past by B-factories \cite{Babar,Belle}, they were the most stringent direct constraints on a light scalar dark boson. 
Their exclusion limits on the coupling (i.e. mixing angle) between the Higgs and the dark boson field lie between $7 \times 10^{-4}$ and $5 \times 10^{-3}$, with the most sensitive region just below the $J/\psi$ threshold \cite{Current_limit}.

This search is performed with the full Run I dataset collected with the LHCb detector corresponding to an integrated luminosity of $3.0 \mbox{ fb}^{-1}$.

\section{Selection and strategy}

Depending on the strength of the mixing with the Higgs boson and its mass, the particle $\chi$ can decay in a secondary vertex, displaced from the $B^0 \rightarrow K^{*0} \chi$ decay vertex. 
In order to increase the sensitivity, two regions of reconstructed di-muon lifetime, $\tau (\mu^+ \mu^- )$, are defined for each $m(\chi)$ considered in the search: a prompt region, $|\tau (\mu^+ \mu^- )| < 3\sigma[\tau (\mu^+ \mu^- )]$, and a displaced region, $\tau (\mu^+ \mu^- ) > 3\sigma [\tau (\mu^+ \mu^- )]$, where $\sigma [\tau (\mu^+ \mu^-)]$ is the lifetime resolution. 
When setting a limit on the branching fraction the two regions are combined as a joint likelihood, $\mathcal{L} = \mathcal{L}^{prompt} \cdot \mathcal{L}^{displaced}$. 
These two regions correspond to the two possible scenarios: the former is sensitive to short lifetime dark boson, it is characterized by high reconstruction efficiency but it is highly contaminated by the irreducible SM background $B^0 \rightarrow K^{*0} \mu^+ \mu^-$; the latter suffers of lower reconstruction efficieny but offers a very clear signature thanks to lower background yields.

A multivariate selection is applied to reduce the background, the uBoost algorithm \cite{uBDT} is employed to ensure that the performance is nearly independent of $m(\chi)$ and $\tau(\chi)$. The inputs to the algorithm include $B^0$ transverse momentum, various topological features of the decay, the muon identification quality, and isolation criteria.
Only candidates with invariant mass $m(B^0)$ within 50 MeV of the known $B^0$ mass are selected.
Then, the reconstructed $m(B^0)$ is constrained to its known value to improve the resolution of the dimuon mass, that results to be less than 8 MeV over the entire $m(\mu^+ \mu^-)$ range, and as small as $2\mbox{ MeV}$ below 220 MeV.

The strategy described in Ref. \cite{strategy_Williams} is adopted: the $m(\mu^+ \mu^-)$ distribution is scanned for an excess of $\chi$ signal candidates over the expected background.
Since all the theoretical models predict the dark boson $\chi$ to have negligible width compared to the detector resolution, the signal window is entirely determinated by the di-muon mass resolution and is defined to be $\pm 2 \sigma[m(\mu^+ \mu^-)]$ around the tested mass.
The step sizes in $m(\chi)$ are $\sigma[m(\mu^+ \mu^- )]/2$. 
In order to avoid experimenter bias, all aspects of the search are fixed without examining the selected $B^0 \rightarrow K^{*0} \chi$ candidates. 

Narrow resonances are vetoed by excluding the regions near the $\omega$, $\phi$, $J/\psi$, $\psi(2S)$ and $\psi(3770)$ resonances. These regions are removed in both the prompt and displaced samples.

\begin{figure}[h!]
	\centering
	\includegraphics[width=13cm]{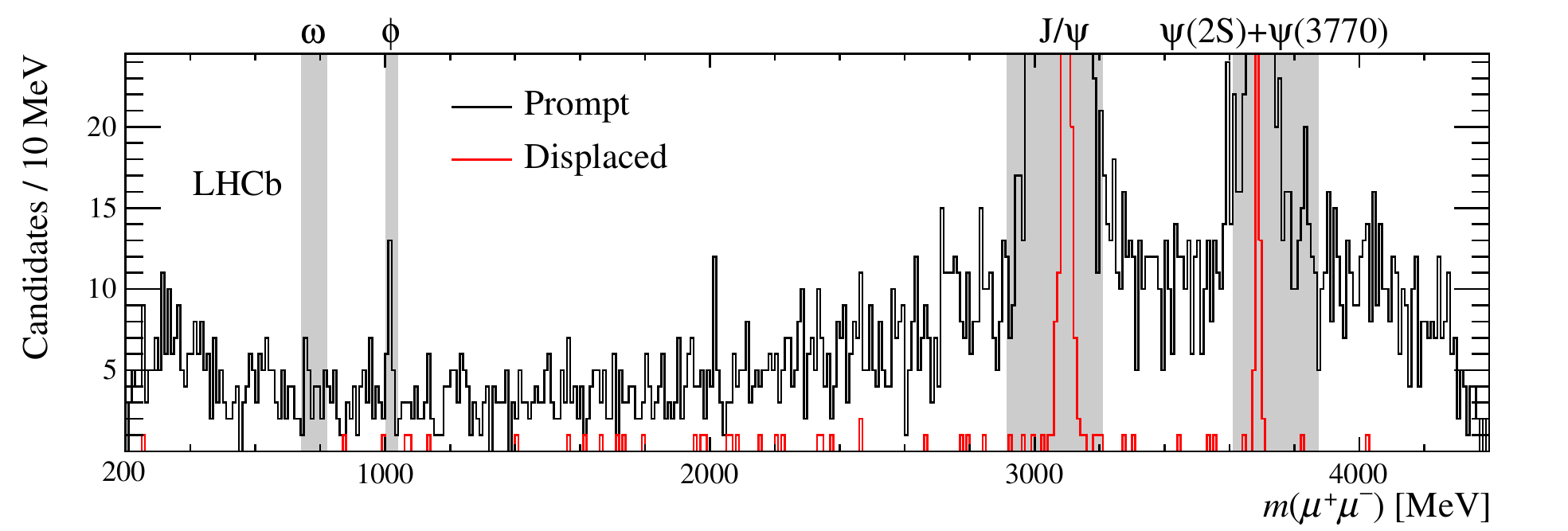}
	\caption{Distribution of $m(\mu^+ \mu^- )$ in the (black) prompt and (red) displaced regions. The shaded bands denote regions where no search is performed due to (possible) resonance contributions. The $J/\psi$, $\psi(2S)$ and $\psi(3770)$ peaks are suppressed to better display the search region.}
	\label{fig:result}
\end{figure}

\begin{figure}[h!]
	\centering
	\includegraphics[width=13cm]{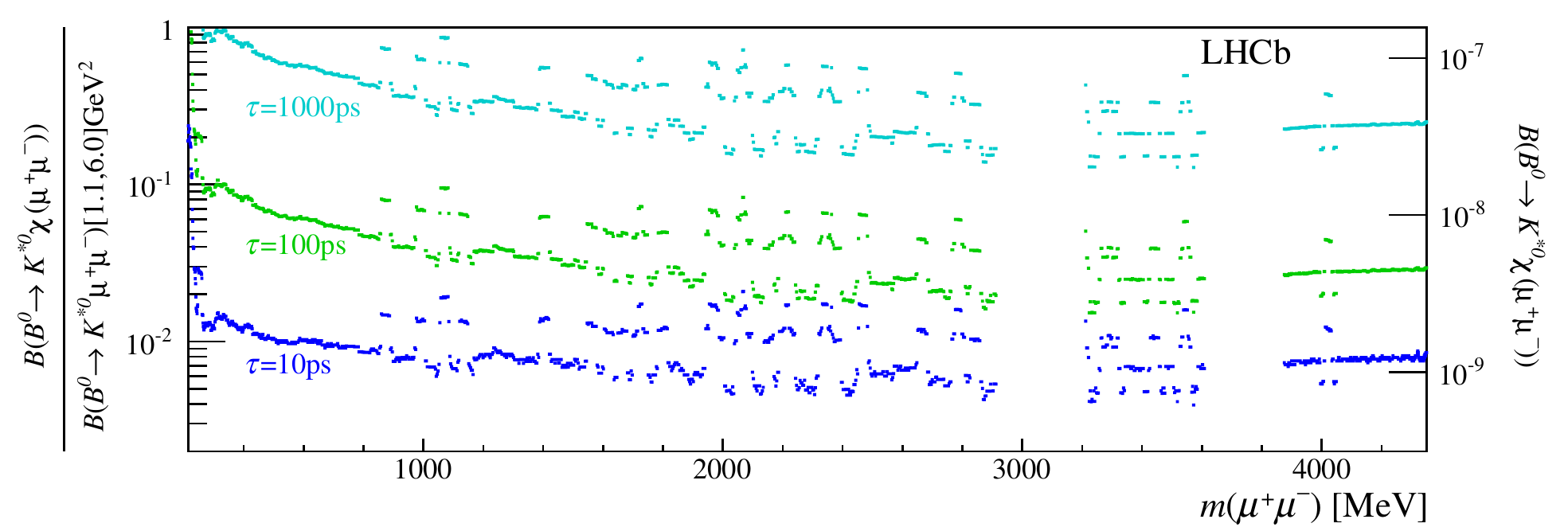}
	\caption{Upper limit on the (left-axis) ratio of branching fractions $\mathcal{B}(B^0 \rightarrow K^{*0} \chi(\mu^+ \mu^- ))/\mathcal{B}(B^0 \rightarrow K^{*0} \mu^+ \mu^- )$, where the $B^0 \rightarrow K^{*0} \mu^+ \mu^-$ decay has $1.1 < m^2 (\mu^+ \mu^- ) < 6.0$ GeV$^2$ and (right-axis) on $\mathcal{B}(B^0 \rightarrow K^{*0} \chi(\mu^+ \mu^- ))$ as a function of the dimuon mass. The limits are given at 95\% confidence level. Limits are presented for three different lifetimes of the dark boson. The sparseness of the data leads to rapid fluctuations in the limits. The relative limits for $\tau < 10$ ps are between $0.005-0.05$ except near $2m(\mu)$.}
	\label{fig:limit_lifetime}
\end{figure}

\section{Results and exclusion limits}

Figure \ref{fig:result} shows the $m(\mu^+ \mu^-)$ distributions for the number of observed candidates in both the prompt and displaced regions.
The observation is consistent with the background only hypothesis with a $p$-value of about 80\%, therefore an upper limit on $\mathcal{B}(B^0 \rightarrow K^{*0} \chi(\rightarrow \mu^+ \mu^-))$ is set.
Figure \ref{fig:limit_lifetime} shows the upper limits both on the absolute branching fraction $\mathcal{B}(B^0 \rightarrow K^{*0} \chi(\mu^+ \mu^-))$ and on the relative ratio to the normalization channel $\mathcal{B}(B^0 \rightarrow K^{*0} \mu^+ \mu^-)$ in the $1.1 < m^2 (\mu^+ \mu^-) < 6.0 $ GeV$^2$ region. Limits are set at the 95\% confidence level (CL) for several values of $\tau (\chi)$.
The limits become less stringent for higher values of $\tau (\chi)$, as the probability of the $\chi$ boson decaying within the LHCb's silicon vertex detector decreases.

\begin{figure}[tb]
	\centering
	\includegraphics[width=7cm]{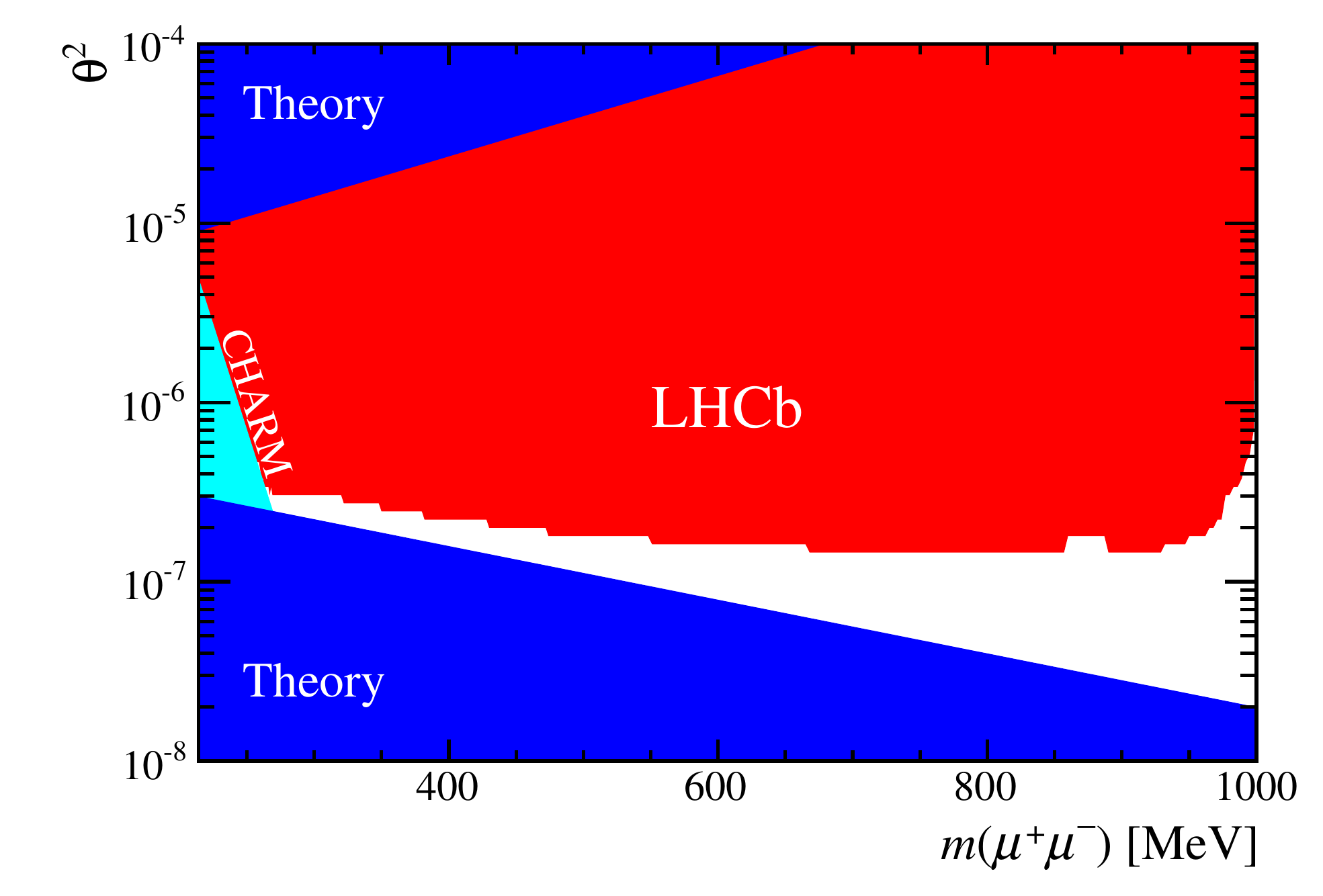}
	\hspace*{2mm}
	\includegraphics[width=7cm]{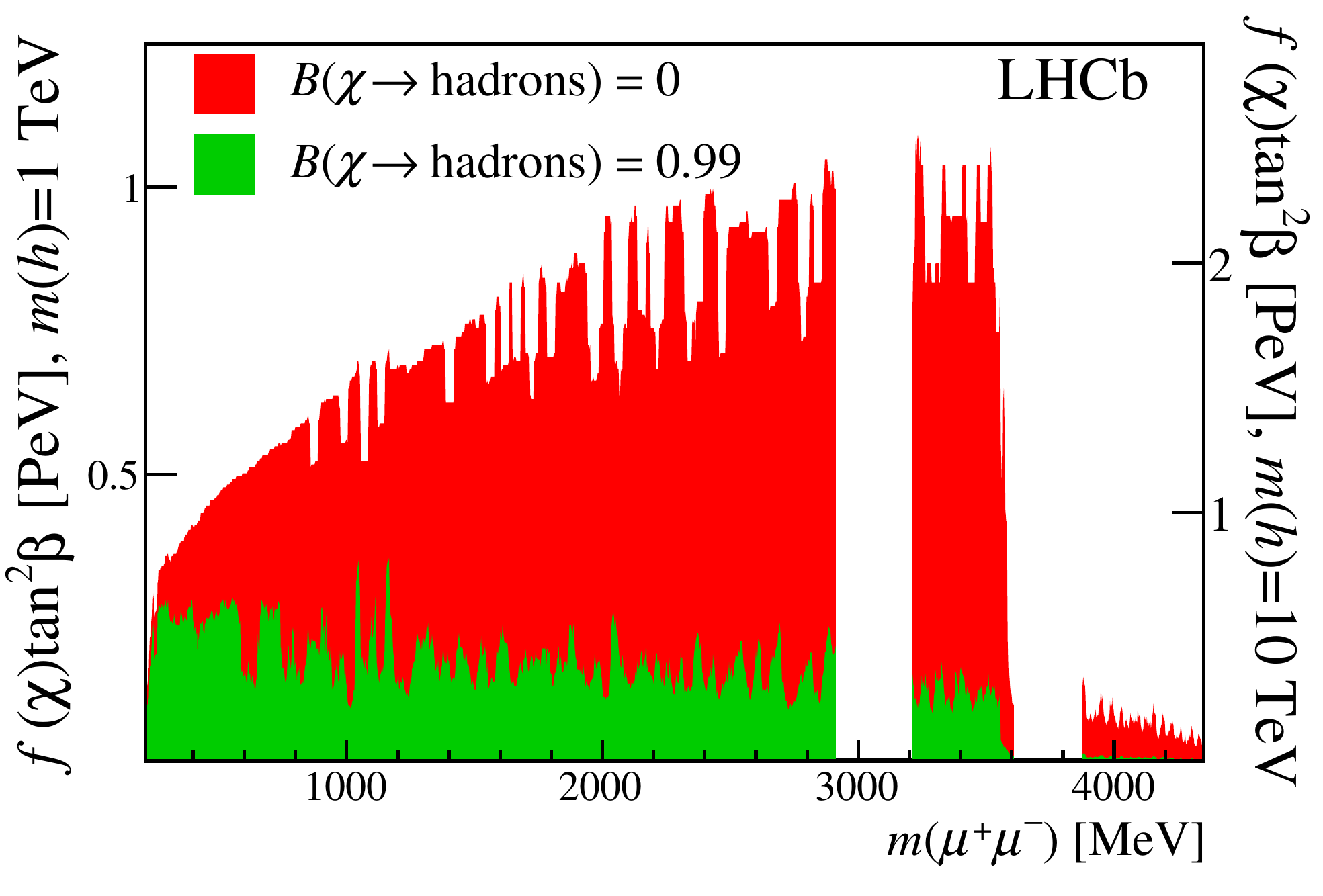}
	\caption{Exclusion regions at 95\% CL: (left) constraints on the inflaton model of Ref. \cite{Bezrukov:Inflaton7keV}; (right) constraints on the axion model of Ref. \cite{Axion}. The regions excluded by the theory \cite{Bezrukov:Inflaton7keV} and by the CHARM experiment \cite{Charm} are also shown.}
	\label{fig:limit_model}
\end{figure}

Figure \ref{fig:limit_model} shows the interpretation of the exclusion limit in term of two benchmark models: the inflaton model of Ref. \cite{Bezrukov:Inflaton7keV}, which only considers $m(\chi) < 1$ GeV, and the axion model of Ref. \cite{Axion}. 
In the first case, constraints are placed on the mixing angle between the Higgs and inflaton fields, $\theta$, which exclude most of the previously allowed region.
For the latter, exclusion regions are set in the limit of large ratio of Higgs-doublet vacuum expectation values, $\tan \beta \gtrsim 3$, for charged-Higgs masses m$(h) = 1$ and 10 TeV.
The branching fraction of the axion into hadrons varies greatly in different models, the results for two extreme cases are shown: $\mathcal{B}(\chi \rightarrow hadrons) = 0$ and 0.99.

\section{Conclusion}

In summary, a search is performed for light scalar dark boson in the decay $B^0 \rightarrow K^{*0} \chi(\rightarrow \mu^+ \mu^-)$ using $pp$-collision data collected at 7 and 8 TeV.
No evidence of signal is observed, and upper limits are placed on $\mathcal{B}(B^0 \rightarrow K^{*0} \chi) \times \mathcal{B}(\chi \rightarrow \mu^+ \mu^-)$. 
This is the most sensitive search to date over the entire accessible mass range and stringent constraints are placed on theories that predict the existence of additional scalar or axial-vector fields.



\end{document}